\documentclass[reprint,groupedaddress,amsmath,amssymb,aps,prl,floatfix,showpacs]{revtex4-1}

\usepackage{graphicx} % Include figure files
\usepackage{dcolumn} % Align table columns on decimal point
\usepackage{bm} % bold math
\usepackage{hyperref} % add hypertext capabilities
\usepackage{times}

\newcommand*{\ket}[1]{\left| #1 \right\rangle}

\newcommand*{\abs}[1]{\left| #1 \right|}

\begin{document}
\title{Phase synchronization between two superradiant lasers} 
\author{Joshua M.~\surname{Weiner}}
 \affiliation{JILA, NIST and Department of Physics, University of Colorado, Boulder, Colorado 80309-0440, USA}
 \email[Corresponding author: ]{weinerjm@jila.colorado.edu}
\author{Kevin C.~\surname{Cox}}
 \affiliation{JILA, NIST and Department of Physics, University of Colorado, Boulder, Colorado 80309-0440, USA}
 \author{Justin G.~\surname{Bohnet}}
 \affiliation{JILA, NIST and Department of Physics, University of Colorado, Boulder, Colorado 80309-0440, USA}
 \author{James K.~\surname{Thompson}}
 \affiliation{JILA, NIST and Department of Physics, University of Colorado, Boulder, Colorado 80309-0440, USA}
\date{\today}
\begin{abstract}
We experimentally demonstrate synchronization between two distinct ensembles of cold atoms undergoing steady state superradiance within a single longitudinal and transverse mode of the same optical cavity. The synchronization process is studied first in terms of the time dynamics of re-synchronization when the phase alignment of the two oscillators is abruptly broken. We also observe the steady state behavior of the lasers as their relative frequency is continuously varied. This system has the potential to realize a non-equilibrium quantum phase transition and could inform future implementations of milliHertz linewidth lasers.
\end{abstract}
\maketitle
Phase synchronization of oscillators is a ubiquitous phenomenon, occurring in physical, chemical, biological, and social systems~\cite{Strogatz2003}. Recent demonstrations of synchronized oscillators at the nano-scale have been reported in mechanical~\cite{SIM2007}, opto-mechanical~\cite{ZWM2012,Bagheri2013}, spintronic~\cite{KPR2005,SPM2013}, and electro-mechanical~\cite{MGV2014} systems. Synchronization dynamics have also been observed in state-of-the-art frequency combs~\cite{Wen2014,DelHaye2014}. Synchronization has been extensively explored with classical models~\cite{Strogatz2000,Pikovsky2003}, and more recently systems have been identified in which quantum noise could contribute to synchronization dynamics~\cite{Lee2013,WNB2014,XTF2014}.  

Synchronized open quantum systems are of fundamental interest in exploring synchronization models~\cite{XTF2014}, associative memories~\cite{Gopalakrishnan2012}, and quantum computing~\cite{VWI2009}. Quantum effects on synchronization are expected in opto-mechanical~\cite{WNB2014,Mari2013,Ludwig2013,Ying2014}, optical~\cite{Lee2013a}, and cold atom~\cite{XTF2014,Lee2014,Lee2013} systems. Here we study synchronization dynamics in the system introduced by Ref.~\cite{XTF2014}: two laser-cooled atomic ensembles that interact through a common cavity mode via steady state superradiant emission. 

The synchronization mechanism of steady state superradiant lasers may enable milliHertz linewidth optical lasers that are highly insensitive to both technical and thermal mirror vibrations~\cite{MYC2009,BCW2012}, an optical analog of the microwave hydrogen maser~\cite{KGR1962}. In such a laser, cavity-mediated interactions combined with repumping-induced dissipation cause the spontaneous synchronization of the phases of the radiating optical dipoles of individual atoms. In the absence of synchronization, the optical dipoles would quickly dephase due to both homogeneous and inhomogeneous broadening, leading to weaker incoherent light emission with a linewidth directly reflecting the width of the broadened atomic transition. Such a narrow optical frequency reference would find a broad range of applications in timekeeping, long-baseline optical interferometry, and precision measurement~\cite{Kessler2012}.  

In this Letter, we study the synchronization of two distinct sub-ensembles of atoms whose relative optical dipole phases can be externally controlled. This allows us to abruptly break the phase alignment between the two collective optical dipoles and watch as synchronization heals the relative phase error. In a second set of experiments, we introduce a continuous source of phase errors (i.e., a frequency offset) between the two ensembles and observe how the competition between the healing rate and the phase error rate leads to a threshold for synchronization to occur. The experimental results are in good agreement with a mean-field model that does not treat quantum noise.

To form the superradiant laser gain medium, we prepare $N = 1.2 \times 10^6$ $^{87}\text{Rb}$ atoms at $20~\mu\text{K}$ within a 1D optical lattice in a high-finesse optical cavity with power decay rate $\kappa = 2\pi\times12~\text{MHz}$ and single-atom cooperativity $C = 5\times10^{-3}$. The lasing transition is a Raman transition from the $\ket{\uparrow} \equiv \ket{5^2 S_{1/2},F=2,m_F=0}$ to $\ket{\downarrow} \equiv \ket{5^2 S_{1/2},F=1,m_F=0}$ ground hyperfine states. In a dressed-state picture, the effective atomic transition frequency is the frequency of the spontaneously emitted Raman photon. The transition frequency is controlled by Raman dressing lasers applied transverse to the cavity axis and tuned 1.3 GHz blue of the $\ket{\uparrow}$ to $\ket{\text{i}} \equiv \ket{5^2 P_{3/2},F=2}$ transition. Repumping from $\ket{\downarrow}$ back to $\ket{\uparrow}$ at single-atom rate $W$ is achieved by applying additional lasers transverse to the cavity mode. The repump lasers are not phase matched with the Raman dressing lasers and their spontaneously scattered photons are not resonant with a cavity mode.

\begin{figure}[!htb]
\includegraphics[width=3.375in]{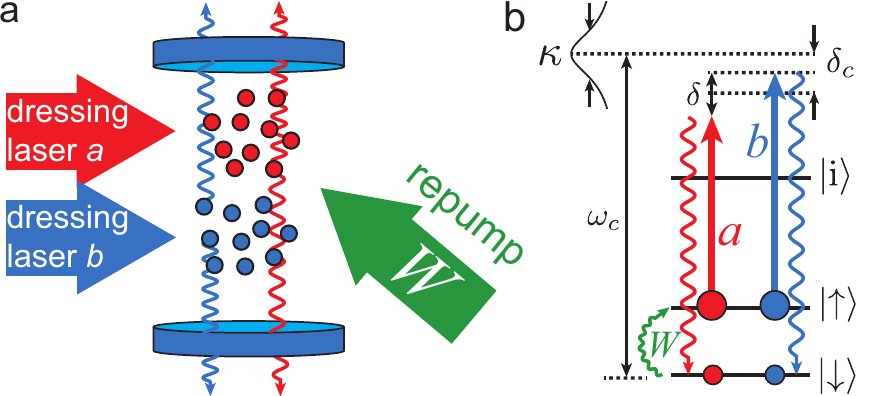}
\caption{(Color online) Experimental diagram and Raman lasing energy levels. (a) Two spatially distinct beams (red, blue) dress an ensemble of laser-cooled atoms inside an optical cavity, defining the two superradiant ensembles $a$ and $b$. Repumping beams (green) are also applied transverse to the cavity. (b) Dressing beams induce Raman decay from $\ket{\uparrow}$ to $\ket{\downarrow}$. Both emitted photon frequencies (wavy lines) are within the linewidth $\kappa$ of a single cavity mode. The repumping laser returns atoms back to $\ket{\uparrow}$ via single-particle repumping at rate $W$.}
\label{fig:fig1}
\end{figure}

To create two spatially separate ensembles with independently controlled optical dipoles, we apply two Raman dressing lasers that address either the upper or lower portions of the total trapped atomic ensemble (Fig.~\ref{fig:fig1}). This provides independent control of the dressing laser phases $\alpha_{a,b}$, angular frequencies $\omega_{a,b}$, and intensities as parameterized by a resonant-Rabi flopping angular frequency $\Omega_{a,b}$ for the $\ket{\uparrow}$ to $\ket{\text{i}}$ transition.  We can independently set the single-atom Raman decay rates $\gamma_{a,b} (\approx 2\pi\times 250~\text{Hz})$ by controlling each laser's intensity. The relative number of atoms $N_{a,b}$ in each ensemble can be controlled by translating the spatial boundary between the dressing lasers along the cavity axis.

Because we utilize Raman transitions for the lasing process, the relevant total optical dipole phases that synchronize are given by $\phi_{a,b} = \eta_{a,b} +\alpha_{a,b}$. Here $\eta_{a,b}$ is the phase associated with the coherence that develops between the ground states $\ket{\uparrow}$ and $\ket{\downarrow}$ in each ensemble. Since the dressing phases are externally controlled parameters, the cavity-mediated interactions drive changes in the ground state coherences $\eta_{a,b}$ to synchronize the optical dipole phases $\phi_{a,b}$.

We first study the dynamics of phase synchronization in the time domain for two ensembles with degenerate frequencies $\delta\equiv \omega_b-\omega_a=0.$  The dressing and repumping lasers are all turned on for 0.1 ms, during which time the two ensembles reach a steady state in which they emit at the same frequency and act as a single synchronized superradiant ensemble with $\phi_a=\phi_b$. An electro-optic crystal is used to quickly jump the phase $\alpha_b$ of the $b$ dressing laser by an amount $\Delta\alpha_b$ in $30$~ns. The timescale of the jump is much faster than the time dynamics of the resynchronization process and effectively creates an instantaneous error in the alignment of the optical phases $\phi_b=\phi_a+\Delta\alpha_b$.

To observe how this phase error heals in time, we allow the system to dynamically evolve for a variable amount of time $T_\text{evol} = 0$~to~$1.5~\mu$s before we rapidly extinguish the other dressing laser, $\Omega_a \rightarrow 0$. Subsequently, only ensemble $b$ radiates into the cavity mode. We infer the change in $\phi_b$ from the difference in the phases $\Delta\psi$ of the emitted light just before the phase jump and just after $T_\text{evol}$.

There are several other physical mechanisms that also affect the observed value $\Delta\psi$ that are not directly related to the synchronization between the optical dipoles. The primary contribution to this background phase shift is population inversion-dependent cavity frequency pulling~\cite{BCW2014}. To remove these less interesting contributions, we measure the light phase difference $\Delta\psi_\pm$ for equal magnitude but opposite sign phase jumps $\pm\Delta\alpha_b$. The computed differential quantity $\Delta\bar{\psi} = (\Delta\psi_+ - \Delta\psi_-)/2$ is insensitive to these background systematic errors. 

\begin{figure}[!htb]
\includegraphics[width=3.375in]{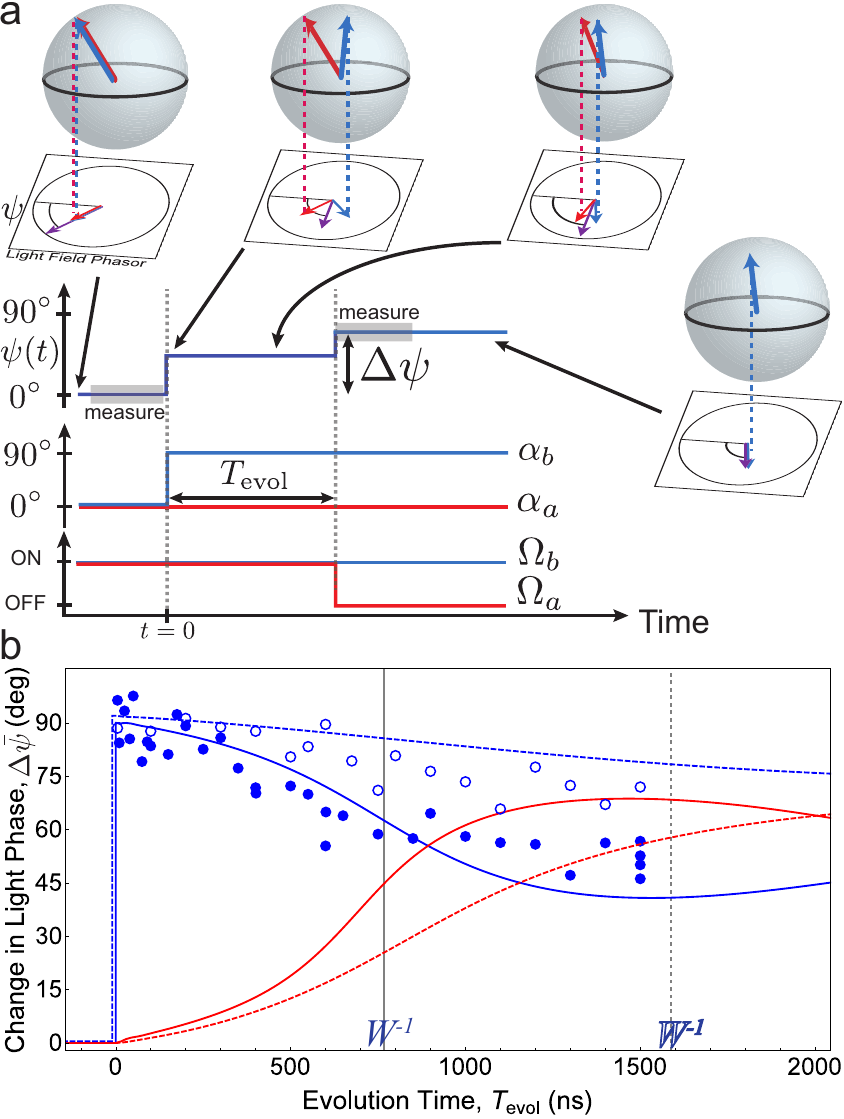}
\caption{(Color) Healing of an instantaneous phase error between optical dipoles. (a) Timing diagram and visualization of atomic Bloch vectors. Before time $t=0$ the two dipoles interact and synchronize. At $t = 0$, dressing laser phase $\alpha_b$ is jumped by $90^\circ$. The ensembles' interaction begins to heal the relative phase error. At $t=T_\text{evol}$, dressing laser $a$ is turned off ($\Omega_a\rightarrow0$) so that only ensemble $b$ radiates into the cavity. The difference $\Delta\bar{\psi}$ in the phases of the radiated light in the grey windows before $t=0$ and after $t=T_\text{evol}$ indicates the change in the optical dipole phase $\Delta \phi_b = \Delta\bar{\psi}$. The upper panels provide cartoon visualizations of phasors representing the radiated fields (red for $a$, blue for $b$, purple for the sum) and Bloch vectors. (b) Light phase change $\Delta\bar{\psi}$ vs. evolution time $T_\text{evol}$. The solid and open points correspond to experiments with dipole ratios $R_d = (1.5,4.0)$ respectively. Vertical solid and dashed lines show the characteristic time scale of the respective single-atom repumping rates for the two data sets $W^{-1} = (0.77, 1.6)~\mu$s corresponding to (solid, open) data. The solid and dashed curves are simulations for the respective data (red for ensemble $a$, blue for $b$).}
\label{fig:fig2}
\end{figure}

The measured quantity $\Delta\bar{\psi}$ as a function of the evolution time $T_\text{evol}$ is shown in Fig.~\ref{fig:fig2}(b). Here the phase jump is $\Delta\alpha_b=90^\circ$, and we see that $\Delta\bar{\psi}$ is also $90^\circ$ near $T_\text{evol}=0$. The phase $\Delta\bar{\psi}$ then relaxes back toward $0^\circ$, settling at an intermediate value such that $\phi_a = \phi_b$. The timescale for relaxation is close to the repumping rate $W^{-1}$, i.e., the characteristic rate at which phase errors are erased, as discussed below.

The equilibrium phase at large $T_\text{evol}$ is mostly determined by the ratio of the relative magnitudes of the optical dipoles of the two ensembles just before the evolution period. The magnitude of each collective dipole is proportional to the number of participating synchronized atoms ($N_{a,b}$) and the emitted electric field per atom ($\propto \sqrt{\gamma_{a,b}}$). The \emph{relative} dipole magnitude is then roughly characterized by $R_d\equiv(N_b\sqrt{\gamma_b})/(N_a\sqrt{\gamma_a})= 1.5$ and $4.0$ for the solid and open data sets in Fig.~\ref{fig:fig2}(b). A simple model for $T_\text{evol}\gg W^{-1}$ and $\Delta\alpha = 90^\circ$ predicts that $\Delta\bar{\psi}$ will relax to $\Delta\bar{\psi}_e \equiv \tan^{-1}(R_d)$. The steady state phase given by the simulation is $\Delta\bar{\psi}_n$. For the data with more balanced populations (solid), the ensembles equally pull each other's optical phases $\phi_{a,b}$ and the light phase relaxes to $\Delta\bar{\psi}=51(3)^\circ$, close to $(\Delta\bar{\psi}_e,\Delta\bar{\psi}_n) = (56^\circ,55^\circ)$. In contrast, in the more imbalanced (open) data, the unobserved $\phi_a$ is pulled more rapdily toward the phase of $\phi_b$, and the phase relaxes toward $\Delta\bar{\psi} = 71(2)^\circ$, while $(\Delta\bar{\psi}_e,\Delta\bar{\psi}_n) = (79^\circ,73^\circ)$, i.e., closer to the phase of ensemble $b$ at $T_\text{evol}=0$. 

Synchronization necessarily implies moving from a state of higher entropy to a state of lower entropy, requiring dissipation into a bath of states that absorb the entropy. In our atom-cavity system, the dominant dissipation mechanism for synchronization is the spontaneously scattered optical pumping light involved in re-exciting the atoms from $\ket{\downarrow}$ to $\ket{\uparrow}$ at rate $W$.  Because our atomic ensemble is optically thin in the direction transverse to the laser cavity, the scattering process for the $i$th atom is not collective and causes single-atom collapse, erasing the relative quantum phase $\phi_i$ in the single atom superposition state: $\cos(\theta_i/2)\ket{\uparrow_i} + e^{\imath \phi_i}\sin(\theta_i/2)\ket{\downarrow_i} \rightarrow \ket{\uparrow_i}$. It is this relative phase $\phi_i$ that encodes the phase of the single-atom dipole and thus the phase of the light $\psi_i = \phi_i+const$ that is radiated by the single oscillator. It is helpful to visualize $\phi_i$ as the azimuthal phase of the single-diople Bloch sphere and the angle $\theta_i$ as a polar angle. Most importantly, the quantum collapse serves to erase any relative phase error $\Delta \phi_i = \phi_i-\phi_\text{avg}$ that had accumulated in time between the individual atom's optical dipole and an appropriately defined average of the phases of all of the optical dipoles of participating atoms $\phi_\text{avg}$.

The total cavity field is the sum of the optical fields radiated by each atom, with a resulting phase $\psi_\text{avg} = \phi_\text{avg}$. This cavity field aligns the optical dipole phase of a newly repumped atom to $\phi_\text{avg}$. The combination of realignment to the average, accrual of phase errors, and erasure of phase errors is the physical origin of the quantum synchronization process.

Another dissipation channel in the atom-cavity system is the emission of photons from the cavity mode through the mirrors. However, this channel only provides collective information to the environment and should not erase single-atom phase errors. Detection of a photon exiting the cavity indicates that one atom has made a transition from $\ket{\uparrow}$ to $\ket{\downarrow}$, but it does not indicate \emph{which} atom made the transition. Still, the fast dissipation ($\kappa\gg \gamma_\perp$) ensures that the cavity field adiabatically follows the total optical dipole moment of the atoms such that the phase of the optical dipoles $\phi_\text{avg}$ can be determined directly from the phase of the emitted light $\psi_\text{avg}$.

\begin{figure}[!htb]
\includegraphics[width=3.375in]{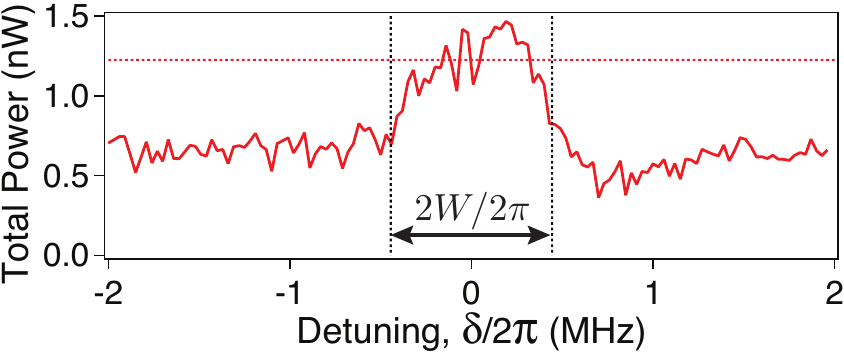}
\caption{(Color online) Total power output vs. detuning for the data shown in Fig.~\ref{fig:fig4}(a). Vertical dashed lines are at the repumping rate $\pm W/2\pi$. Horizontal dashed line is the predicted maximum synchronized output power.}
\label{fig:fig3}
\end{figure}

\begin{figure}[!htb]
\includegraphics[width=3.375in]{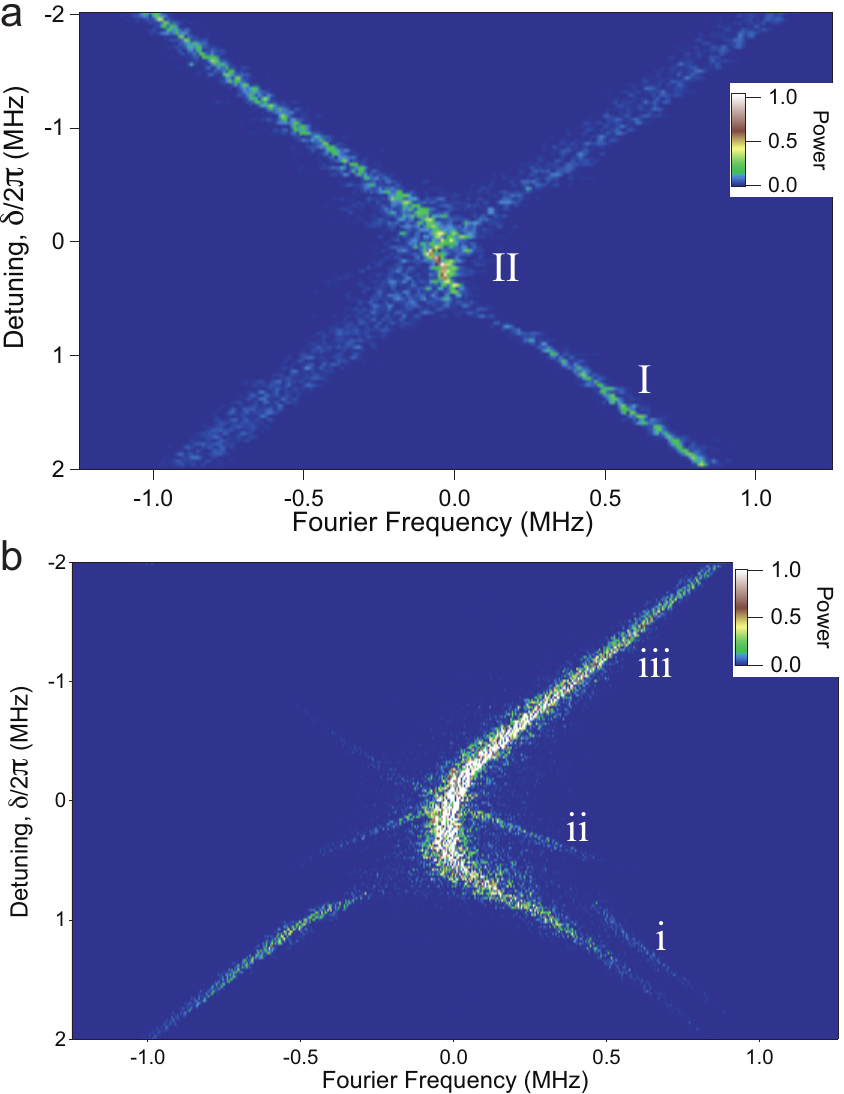}
\caption{(Color) Spectrograms of light emitted from two superradiant ensembles. Vertical axis is the detuning of dressing lasers $\delta$ and horizontal axis is the Fourier frequency of each power spectrum. The power (color scale) is normalized to the maximum power across the entire spectrogram. (a) Each power spectrum displayed here represents the mean of 5 power spectra at each $\delta$. Collective dipoles are roughly balanced with $N_a/N_b = 0.6$ and $\gamma_a/\gamma_b = 0.8$ (b) Asymmetric operating conditions, $N_a/N_b = 1.1$ and $\gamma_a/\gamma_b = 1.6$.}
\label{fig:fig4}
\end{figure}

We next consider the case in which a continuous source of phase error is introduced between the two ensembles. This is equivalent to detuning the dressing laser frequencies. As $\delta$ deviates from zero, the total power emitted by the two ensembles decreases as shown in Fig.~\ref{fig:fig3}. For $\left|\delta\right| > W$ the total output power is roughly constant. At the transition point, the two ensembles largely behave independently, emitting at their respective natural lasing frequencies. The characteristic frequency scale is set by $W$ since any relative phase accumulated between the ensembles is reset by repumping. The observed maximum synchronized power output is a factor of 2.2(1) greater than the unsynchronized power output, while we predict a factor of $1.8(2)$. This estimate is based on the quenching behavior of the output power with repumping rate that accounts for changes in population inversion of each ensemble~\cite{BCW2012,Meiser2010}. The asymmetry of the total power for positive and negative $\delta$ is also reflected in the asymmetric behavior in the spectra of Fig.~\ref{fig:fig4} and discussed below.

We can observe the transition from synchronized to unsynchronized behavior in the frequency domain by making heterodyne measurements of the light emitted from the cavity.  In the spectrograms of Fig.~\ref{fig:fig4}, each row is a frequency spectrum of emitted light from the cavity, with brighter colors indicating higher power. Each power spectrum is calculated from $80~\mu\text{s}$ of the time record. The two-dimensional power spectrum is created by repeating the measurement at a series of different detunings $\delta$, with values shown along the vertical axis.

For $\left|\delta\right| \gg W$, the two ensembles of atoms emit at frequencies very close to the unperturbed Raman transition frequencies. As $\abs{\delta}$ decreases, the emission frequencies are pulled toward each other as the rate of relative phase error introduction $\delta$ nears the error erasure rate $W$. We note that we do not observe nor expect a region of repulsive synchronization that appears when injection locking a single superradiant ensemble to an externally applied drive~\cite{CWT14}. For $\left|\delta\right| \lesssim W$, the erasure of phase errors dominates and the two ensembles radiate at a single frequency.  

The observed spectrum qualitatively agrees with the same mean-field model used to describe the time dynamics of resynchronization, exhibiting a hyperbolic-like approach (Region I) to the synchronized state (Region II). However, there is significant asymmetry in the power spectrum. Part of this asymmetry arises from a finite detuning of the average Raman transition frequency from resonance with the cavity resonance frequency by amount $\delta_c = -2\pi\times 4~\text{MHz} \approx \kappa/3$, an operating condition favorable for suppressing relaxation oscillations~\cite{BCW2012a,BCW2014} yet one that introduces an imbalance in the coupling to the cavity between ensembles.  Other causes of asymmetry are imbalances in the optical dipole magnitudes (both $N$ and $\gamma$) for the data in Fig.~\ref{fig:fig4}. Numerical modeling indicates that the effects of these small asymmetries are magnified by the interaction between the ensembles. 

We also show in Fig.~\ref{fig:fig4}(b) that many different behaviors can be observed depending on the operating parameters. This data shows a significant asymmetry in the emitted power (iii) from each ensemble for $\delta > 0$ and $\delta < 0$. Many of these behaviors are observed in numerical mean-field models of our system, but other features, indicated in Fig.~\ref{fig:fig4}(b) are not: (i) the parallel-running frequency component in the lower right hand quadrant, (ii) the extra frequency components at $\pm \delta/2\pi$, and the asymmetry in the observed linewidth of the two emission peaks of both Fig.~\ref{fig:fig4}(a) and (b). The fractional power in each sideband (ii) is small, $<8\%$ of the total power in each spectrum.

In prior studies, linewidth broadening was seen to arise from an inversion-dependent frequency-pulling mechanism that here would cause a common broadening of both peaks~\cite{BCW2012,BCW2014}. Attempts to identify other classical mechanisms for the asymmetric broadening have been unsuccessful and the broadening phenomenon remains an interesting topic for future theoretical and experimental study, with the intriguing possibility that this is a fundamental quantum noise effect~\cite{XTF2014}.

This work emphasizes the key role that dissipation via repumping plays in erasing the phase errors that are generated during the synchronization process. This work may apply to future technologically relevant implementations of superradiant ensembles that would produce optically narrow light~\cite{MYC2009,BCW2012}. For instance, two superradiant lasers that operate on atomic transitions with opposite sensitivity to magnetic fields could use synchronization to cancel frequency transition noise as must be done sequentially in passive optical lattice clocks~\cite{Bloom2014}. Also, this work points toward a coupled atom-cavity system for exploring quantum noise in phase transition models~\cite{XTF2014}, using atomic synchronization for enhanced Ramsey spectroscopy~\cite{Xu2014}, and overcoming the Dick effect by transferring coherence between atomic clocks~\cite{RL2013,BS2013}. 
\begin{acknowledgments}
The authors acknowledge early contributions by Matthew A. Norcia. All authors acknowledge financial support from DARPA QuASAR, ARO, NSF PFC, and NIST. K.C.C. acknowledges support from NDSEG. This work is supported by the National Science Foundation under Grant Number 1125844.
\end{acknowledgments}
\bibliographystyle{apsrev4-1}
\bibliography{main}
\end{document}